\documentclass[11pt]{article}

\usepackage{graphicx}
\usepackage[reqno,tbtags]{amsmath}
\usepackage{multirow}
\usepackage{cite}
\usepackage{epsfig}

\allowdisplaybreaks

\newcommand\pubnumber{ PITHA 08/21  \\
                       SFB/CPP-08-63  \\
                       TTP08-39  }

\newcommand\pubdate{\today}

\def\csuma{Institut f\"ur Theoretische Physik E, RWTH Aachen University,\\
           52056 Aachen, Germany}
\def\csumb{Dipartimento di Fisica Teorica, Universit\`a di Torino, Italy\\
           INFN, Sezione di Torino, Italy}
\def\csumc{Physics Department, Brookhaven National Laboratory,\\
           Upton, NY 11973, USA}
\def\csumd{Institut f\"ur Theoretische Teilchenphysik, Universit\"at Karlsruhe,\\
           76128 Karlsruhe, Germany}

\def\Title#1{\begin{center}{\Large \bf #1}\end{center}}

\def\Author#1{\begin{center}{\sc #1}\end{center}}

\def\Address#1{\vspace{-0.5cm}\begin{center}{\it #1}\end{center}}

\newcommand\pubblock{\rightline{\begin{tabular}{l}\pubnumber\\
                     \pubdate\\\end{tabular}}}

\newenvironment{Abstract}{\begin{quotation}}{\end{quotation}}

\def\Acknowledgments{\bigskip\bigskip\begin{center}
                     \large\bf Acknowledgments\end{center}}

\def\email#1{\footnote{#1}}

\makeatletter

\def\section{\@startsection{section}{0}{\z@}{5.5ex plus .5ex minus
                     1.5ex}{2.3ex plus .2ex}{\large\bf}}

\def\subsection{\@startsection{subsection}{1}{\z@}{3.5ex plus .5ex minus
                          1.5ex}{1.3ex plus .2ex}{\normalsize\bf}}

\def\subsubsection{\@startsection{subsubsection}{2}{\z@}{-3.5ex plus
                                      -1ex minus  -.2ex}{2.3ex plus .2ex}{\normalsize\sl}}

\input Input_rosetta.sty

\begin{document}

\begin{titlepage}
  \pubblock
  \vfill
  \def\thefootnote{\fnsymbol{footnote}}
  \Title{Two-Loop Threshold Singularities,\\[3mm]
         Unstable Particles and Complex Masses 
  \footnote[9]{Work supported by MIUR under contract 2001023713$\_$006, 
               by the European Community's Marie Curie Research 
               Training Network {\it Tools and Precision Calculations for 
               Physics Discoveries 
               at Colliders} under contract MRTN-CT-2006-035505, by the U.S. 
               Department of Energy under contract No. DE-AC02-98CH10886 and
               by the Deutsche Forschungsgemeinschaft through 
               Sonderforschungsbereich/Transregio 9 
               {\it Computergest\"utzte Theoretische Teilchenphysik}.
               The authors thank the Galileo Galilei institute for Theoretical Physics
               for hospitality and the INFN for partial support during the completion of
               this work.}}
\vfill
\Author{Stefano Actis       \email{actis@physik.rwth-aachen.de}}        \Address{\csuma}
\Author{Giampiero Passarino \email{giampiero@to.infn.it}}               \Address{\csumb}
\Author{Christian Sturm     \email{sturm@bnl.gov}}            \Address{\csumc}
\Author{Sandro Uccirati     \email{uccirati@particle.uni-karlsruhe.de}} \Address{\csumd}
\vfill
\begin{Abstract}
\noindent
  The effect of threshold singularities induced by unstable
  particles on two-loop observables is investigated and it is shown how to cure 
  them working in the complex-mass scheme. The impact on radiative corrections 
  around thresholds is thoroughly analyzed and shown to be relevant for two 
  selected LHC and ILC applications: Higgs production via gluon fusion and decay 
  into two photons at two loops in the Standard Model. Concerning Higgs production, 
  it is essential to understand possible sources of large corrections in addition 
  to the well-known QCD effects. It is shown that NLO electroweak corrections can 
  incongruently reach a $10\, \%$ level around the $WW$ vector-boson threshold 
  without a complete implementation of the complex-mass scheme in the two-loop 
  calculation.
\end{Abstract}
\vfill
\begin{center}
\end{center}
\end{titlepage}
\def\thefootnote{\arabic{footnote}}
\setcounter{footnote}{0}
\clearpage
\setcounter{page}{1}
\section{Introduction}
The computation of higher order corrections to multi-scale processes 
is plagued by the presence of unstable particles in loop integrals. 
Formally, a clean description would require a Dyson resummation of 
self-energy insertions in order to preserve unitarity~\cite{Veltman:1963th}; 
in the context of the Standard Model and its extensions, however,
the consequent mixing of perturbative orders clearly compromises 
gauge invariance. In addition to the unitarity issue, a practical problem 
is represented by the appearance of unphysical threshold singularities at the 
amplitude level for physical observables. 

In this letter we focus on two-loop electroweak corrections to Standard 
Model Higgs production through gluon fusion, $gg\to H$, and decay into 
two photons, $H \to \gamma\gamma$ (details of the calculation will
be given in a forthcoming paper~\cite{newLong}). We discuss the singular behavior 
of the amplitudes around the normal thresholds induced by internal unstable 
particles, directly related to two-particle unitarity cuts. 
Concerning the  $H \to \gamma\gamma$ case, note that threshold effects have been 
analyzed in Ref.~\cite{Drees:1989du}, with special emphasis on the presence of 
bound states, which are not the subject of our study.

The singular behavior can be cured trading the real masses for unstable 
particles, used as experimental input data, with the associated complex poles. 
As shown by the authors of Ref.~\cite{Denner:2005fg}, the replacement has to 
be performed also at the level of the couplings, leading to the so-called 
complex-mass scheme. Consequently, the one- and two-loop integrals needed 
for the computation have to be evaluated with complex internal arguments.

A minimal implementation of the complex-mass scheme for two-loop electroweak 
corrections to the $H \to \gamma\gamma$ process has been realized in 
Ref.~\cite{Passarino:2007fp}. The two-loop amplitude is splitted in gauge-invariant 
divergent and non-divergent terms, and complex masses are introduced in the 
divergent part. This solution is formally satisfactory, since the amplitude is 
finite also at thresholds. Furthermore, as a consequence of the cancellation 
mechanism taking place among divergent contributions, this scheme does not require 
any analytic continuation of two-point functions connected to mass renormalization 
to the second Riemann sheet.

As a drawback, artificially large numerical effects arise around normal 
thresholds. The issue is relevant when we consider Higgs production 
through gluon fusion at the LHC, where a measurement of the Higgs mass
can be performed at the per-mille level~\cite{HiggsMass}. The statement that 
next-to-leading order electroweak corrections are known~\cite{:2008uu} does not 
appear to be fully adequate, given the possible occurrence of large threshold 
effects of about $10\, \%$ with respect to the leading order result. 

In this letter we show that threshold effects for $gg\to H$ and $H\to 
\gamma \gamma$ are well under control in our improved 
calculational scheme~\cite{newLong}, where a complete implementation of the 
complex-mass scheme at two loops is performed along the lines of 
Ref.~\cite{Actis:2006rc}. 
For the analysis in hadron-hadron collisions we refer to 
Ref.~\cite{newShort}.
\section{Radiative corrections with unstable particles}
In this section we briefly summarize aspects of selected solutions
for dealing with unstable particles in tree and one-loop calculations;
next, we analyze the salient features of a two-loop computation. 
In \sect{sec:sing} we will study where the numerical impact of complex masses 
is most relevant at the two-loop level by looking into the presence of unphysical
infinities and cusps in two-loop amplitudes. 

At tree level, if the external legs of a given amplitude are divided into 
two disjoint subsets and if the total quantum numbers of each subset allow for
the exchange of a known particle of mass $M$, then the amplitude has a pole 
satisfying $p^2 = M^2$, where $p$ is the four-momentum of the exchanged 
particle.
In leading order (LO) calculations, the masses of these particles, like $W$ and $Z$ bosons, are
replaced by the location of the poles in the complex $p^2$ plane.
However, the principle of gauge invariance must not be violated: 
in particular, Ward-Slavnov-Taylor (hereafter WST) identities~\cite{Ward:1950xp}  have to be 
preserved, otherwise theoretical uncertainties may get out of control. 
The incorporation of finite-width effects in the theoretical predictions for Lep2 
processes was a typical example and, at the time, it was argued that the preferable
scheme consists in resummation of fermion one-loop corrections to vector-boson 
propagators and inclusion of all remaining fermion one-loop corrections, in 
particular those to the Yang-Mills vertices~\cite{Beenakker:1996kn}.

A possible solution at the next-to-leading order (NLO) level consists in replacing everywhere the squared real 
masses ($m^2$) with the complex poles ($s_\ssP$), couplings included; this is known in the literature 
as complex-mass scheme~\cite{Denner:2005fg}.
Since WST identities are algebraic relations satisfied separately by real
and imaginary parts, one starts from WST identities with real masses,
satisfied at any given order, and replaces everywhere $m^2 \to s_{\ssP}$ without
violating the invariance. 

In turns, this scheme violates unitarity: one cannot identify the two 
sides of any cut diagram with $T$ and $T^{\dagger}$ respectively (the
transition matrix $T$ is defined in terms of the $S$ matrix as
$S= 1+i\,T$). 
To summarize, the analytical structure of the $S$ matrix is correctly 
reproduced when we use propagator factors $s - s_{\ssP}$, where $s$ is a
generic invariant, but unitarity of $S$ requires more, a dressed 
propagator~\cite{Actis:2006ra,Actis:2006rc}.
However, we expect that unitarity-violating terms are of higher order; in
principle, the violating terms should not be enhanced because WST identities are
preserved.

Another drawback of the scheme is that all propagators for unstable particles
will have the same functional form both in the time-like and 
space-like regions, while, for a dressed propagator, the presence of a pole
on the second Riemann sheet does not change the real character of the function
if we are in a $t$ channel.

Typical examples of one-loop calculations that require the introduction of complex
poles are those for processes, like $e^+e^- \to 4\,$f, where part of the amplitude
(the so-called signal) factorizes into production$\,\otimes\,$decay of
one or more particles; or processes involving off-shell $W\,$-pair production,
leading to the so-called Coulomb singularity. Another approach in this
context is represented by the use of effective field theory methods
(see recent applications in Ref.~\cite{Beneke:2007zg}).

In some sense the complex-mass scheme becomes more appealing when we go beyond one 
loop, as described in detail in Ref.~\cite{Actis:2006rc}, where a recipe was 
designed to derive loop amplitudes out of a skeleton expansion. 
Let $m^2$ be the squared bare mass, $s_{\ssP}$ the corresponding complex pole 
and $\Sigma$ the corresponding self-energy: to the requested order we replace 
everywhere $m^2$ with $s_{\ssP} + \Sigma(s_{\ssP})$, which is real by construction.
If only one loop is needed, then $m^2 \to s_{\ssP}$ everywhere (therefore
justifying the name {\em complex mass}).

Note that the on-shell mass is related to the zero of the real part of the 
inverse propagator; beyond one loop this would show a clash with gauge invariance,
since only the complex poles do not depend on gauge parameters to all orders.
As a consequence, renormalization equations change their structure.

Furthermore, there is also a change of perspective with respect to one-loop 
calculations. 
There one considered the on-shell masses as input parameters independent of 
complex poles and derived the latter in terms of the 
former~\cite{Beenakker:1996kn}. 
Here the situation changes: renormalization equations are written for
real renormalized parameters and solved in terms of (among other things) 
experimental complex poles; the latter have to be reconstructed from on-shell 
pseudo-observables. 

Having described the general setup for a gauge-invariant formulation of unstable
particles at the multi-loop level, we want to understand where, in a two-loop
calculation, the numerical impact of complex masses is most relevant. 
\section{Two-loop amplitudes and normal thresholds}
\label{sec:sing}
In this section we explore the singular behavior of massive two-loop
amplitudes around normal thresholds, investigating the origin 
of square-root and logarithmic singularities. It is worth noting that 
pseudo-thresholds are always outside the physical region.

\begin{figure}[ht]
    \begin{center}
      \includegraphics[scale=0.6]{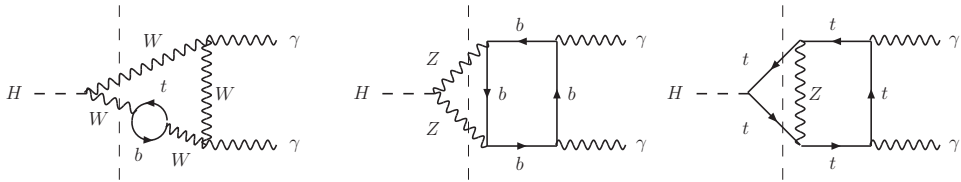}
    \end{center}
    \caption[]{Sample cut diagrams for $H \to \gamma \gamma$ showing normal
      thresholds for $\mh=2\mw,2\mz,2M_t$.}
    \label{Cuts}
  \end{figure}

Normal thresholds are directly related to unitarity cuts as illustrated in 
\fig{Cuts}, and correspond to the leading Landau singularities~\cite{Landau:1959fi} 
of self-energy diagrams. When diagrams with more than two external legs are 
present, normal thresholds show up as sub-leading singularities; this can be easily 
understood observing that all diagrams in \fig{Cuts} are reduced to 
self-energy configurations after shrinking a line which does not intersect 
any cut to a point.

An interesting question is the role played by the leading singularity (the
so-called anomalous threshold)~\cite{Goria:2008ny}: already at one loop, there are 
cases where a non-integrable singularity associated with the leading Landau 
singularity requires the introduction of complex masses~\cite{Boudjema:2008zn}. 
Concerning the processes $H\to \gamma \gamma$ and $gg \to H$, we have
verified that the leading Landau singularities of all two-loop vertex diagrams lie 
outside the physical region, and we will drop this issue in the following.

For our discussion, it is useful to decompose
the $H\to \gamma \gamma$ amplitude as
  \bq\label{eq:deco}
  {\cal A}= 
  \lpar \srt \, \gf \mws \rpar^{1\slash 2} 
  \frac{\alpha}{2\pi}
  \lpar {\cal A}^{\rm LO} + \frac{\gf \mws}{2\srt\, \pi^2} {\cal A}^{\rm NLO}\rpar,
  \eq
where $\gf$ is the Fermi-coupling constant, $\mw$ is the mass of the $W$
boson and $\alpha$ is the fine-structure constant; ${\cal A}^{\rm LO}$
and ${\cal A}^{\rm NLO}$ denote the leading order and next-to-leading
order amplitudes. Furthermore, ${\cal A}^{\rm NLO}$ can be written
as
  \bq\label{eq:trivialsplit}
  {\cal A}^{\rm NLO}= {\cal A}^{\rm 2L} +  
                      {\cal A}^{\rm REN} +
                      {\cal A}^{\rm WFR},
  \eq
where ${\cal A}^{\rm 2L}$ is given by the sum of all pure two-loop diagrams,
${\cal A}^{\rm WFR}$ follows from the inclusion of the one-loop Higgs
wave-function renormalization (WFR) factor and ${\cal A}^{\rm REN}$
stems from one-loop renormalization of the masses and the Fermi coupling.
Note that the U(1) Ward identity forces the electromagnetic coupling to 
go unrenormalized once the external on-shell photons are provided with their
WFR factors.
For electroweak corrections, the amplitude for
$gg \to H$ is given by \eqn{eq:deco}, with 
$\alpha$ replaced by the strong-coupling constant $\alpha_\ssS(\mu_\ssR^2)$ at the 
renormalization scale $\mu_\ssR$.
\subsection{Square-root singularities}
A square-root singularity is represented by a term containing a single inverse
power of the threshold factor $\beta_i$,
  \bq\label{eq:beta}
    \beta_i = \sqrt{1 - 4\,M_i^2 \slash \mhs}, \qquad \text{with} 
\qquad M_i=\mw,\mz,M_t.
  \eq

In Ref.~\cite{Passarino:2007fp} it has been shown that square-root singularities
are related to: i) derivatives of two-point one-loop functions, associated
with Higgs WFR; ii) derivatives of three-point one-loop functions, 
generated by mass renormalization; iii) genuine irreducible two-loop diagrams 
containing a one-loop self-energy insertion. Therefore, in general, all three 
terms of \eqn{eq:trivialsplit} can show a $\beta_i^{-1}$ behavior.

\begin{figure}[ht]
    \begin{center}
      \includegraphics[scale=0.6]{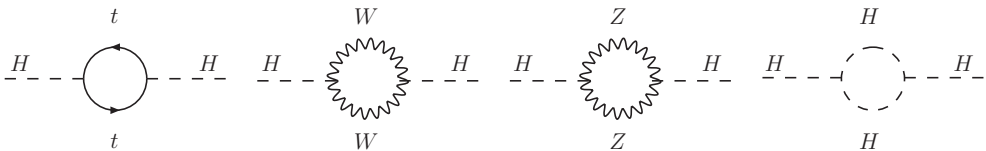}
    \end{center}
    \caption[]{Sample diagrams for the Higgs wave-function renormalization factor 
    at one loop.}
    \label{Higgs}
  \end{figure}

Concerning the Higgs WFR factor at one loop, we deal with the four mass 
patterns shown in \fig{Higgs}; note that tadpole diagrams do not affect
the threshold behavior. For the top-quark diagram, the coefficient of the derivative
of the two-point one-loop function contains a positive power of the threshold
factor $\beta_t$; in other words, this diagram is $\beta_t$-protected at threshold.
Consequently, ${\cal A}^{\rm WFR}$ of \eqn{eq:trivialsplit} contains, for both 
processes $H\to \gamma \gamma$ and $gg \to H$, square-root singularities only
at the $2\, \mw$ and $2\, \mz$ thresholds.

\begin{figure}[ht]
    \begin{center}
      \includegraphics[scale=0.6]{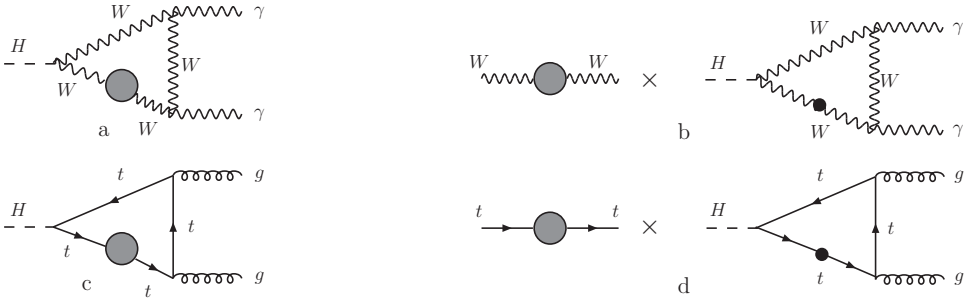}
    \end{center}
    \caption[]{Two-loop and mass-renormalization diagrams relevant for the
      analysis of square-root singularities. Gray circles represent the sum
      of all one-loop two-point diagrams; black dots denote a derivative.}
    \label{4diags}
  \end{figure}

We consider now genuine two-loop diagrams containing a self-energy insertion;
they naturally join terms induced by one-loop mass renormalization as shown
in \fig{4diags}, where bosonic and fermionic diagrams are illustrated. The bosonic
component is obviously peculiar of the $H\to \gamma \gamma$ decay; in addition,
we observe that only charged bosonic diagrams are present, because
of the nature of the triple non-abelian gauge coupling in the Standard Model.

Fermionic diagrams are $\beta_t$-protected at threshold, and do not require
any special care. The two-loop vertex containing a $W$ self-energy insertion,
and the associated $W$-mass renormalization contribution,
lead instead to a $\beta_\ssW^{-1}$-divergent behavior. 
However, the two-loop irreducible diagram of \fig{4diags}a can be cast 
in a representation where the singular part is completely written in terms of 
the one-loop diagrams of \fig{4diags}b.
Moreover, it is possible to check explicitly that the unphysical 
$\beta_\ssW^{-1}$ behavior, generated by the two-loop diagram of \fig{4diags}a, 
exactly cancels the  $\beta_\ssW^{-1}$ divergency due to one-loop $W$-mass 
renormalization of \fig{4diags}b (performed in the complex-mass setup we are 
going to describe in Section~\ref{sec:CM2}).

This cancellation mechanism corroborates the general
picture of Ref.~\cite{Veltman:1963th}: self-energy insertions signal the presence 
of an unstable particle, and are the consequence of a misleading organization of 
the perturbative expansion; Dyson-resummed propagators should be used and complex 
poles should replace real masses as input data.
The outcome of our analysis, performed at the algebraic level, is straightforward:
at the amplitude level, only the inclusion of the Higgs WFR factor generates 
square-root singularities at the $2 \mw$ and $2 \mz$ thresholds for both processes 
$H\to \gamma \gamma$ and $gg \to H$.
\subsection{Logarithmic singularities}
\begin{figure}[ht]
    \begin{center}
      \includegraphics[scale=0.55]{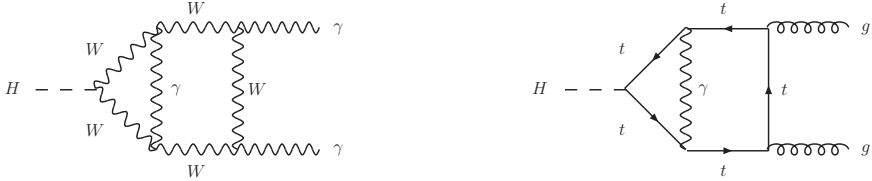}
    \end{center}
    \caption[]{Irreducible two-loop vertex
      diagrams which can generate a logarithmic divergency.}
    \label{TLvertbca}
 \end{figure}

We briefly address here the issue of logarithmic singularities,
given by terms containing a factor $\ln(-\beta_i^2-i 0)$, generated
by the diagrams of \fig{TLvertbca}.
As thoroughly discussed in Ref.~\cite{Passarino:2007fp}, the scalar
configuration associated with the diagrams illustrated in \fig{TLvertbca}
generates a logarithmic singularity. If the massive loop is made of top quarks, 
the scalar integral appears at the amplitude level with a multiplicative
factor $\beta_t^2$, and the logarithmic singularity is $\beta_t^2$-protected at 
threshold. The same consideration is not valid for a $W\,$ loop; here the 
logarithmic singularity can be viewed as the remnant of a Coulomb singularity
in the one-loop sub-diagram.
\vspace{-0.7cm}
\section{Complex masses}
\label{sec:CM2}
A pragmatic gauge-invariant solution to the problem of threshold 
singularities due to unstable particles for the $H\to \gamma \gamma$ 
decay has been introduced and formalized in Ref.~\cite{Passarino:2007fp} (ad hoc
introduction of a width is however common practice in the literature). After 
reviewing the corresponding scheme, termed in the following as minimal 
complex-mass (MCM) scheme, we will discuss the extension to the full 
complex-mass (CM) scheme, looking ahead to precise predictions for 
the production mechanism $g g \to H$.
\subsection{Minimal complex-mass scheme}
\label{sec:MCMscheme}
In the MCM scheme the NLO amplitude of \eqn{eq:deco} and \eqn{eq:trivialsplit}
is decomposed according to
  \bq\label{eq:MCMsplit}
  {\cal A}^{\rm NLO}= \sum_{i=W,Z} \frac{A_{{\rm SR},i}}{\beta_i}
                     + A_{{\rm LOG}} \ln \lpar -\beta_W^2 - i 0 \rpar
                     + A_{{\rm REM}},
  \eq
where square-root- ($A_{{\rm SR},i}$ ) and logarithmic-singular ($A_{{\rm LOG}}$) 
terms have been isolated from the component which is finite for $\beta_i\to 0$ 
($A_{{\rm REM}}$). 
From the discussion of Section~\ref{sec:sing}, we know that $A_{{\rm SR},i}$, 
with $i=W,Z$, is generated for both $H\to \gamma \gamma$ and $g g \to H$ 
by Higgs WFR at one loop; $A_{{\rm LOG}}$ shows up for $H\to \gamma \gamma$ 
only, and is induced by the bosonic diagram of \fig{TLvertbca}.

After proving that all coefficients in \eqn{eq:MCMsplit}, gauge-parameter
independent by construction, satisfy the WST identities, we minimally
modify the amplitude introducing the complex-mass scheme of 
Ref.~\cite{Denner:2005fg} for the divergent terms.
In principle, two steps are required: first, the real masses of the $W$ and $Z$ 
bosons, used as input data, are replaced by the corresponding complex poles
in the threshold factors $\beta_i$, $i=W,Z$, and in the coefficients 
$A_{{\rm SR},i}$ and $A_{{\rm LOG}}$ (also at the level of the couplings); second, 
the real parts of the $W$ and $Z$ self-energies stemming from mass renormalization 
at one loop are traded for the complete self-energies, including imaginary parts. 

In practice, the second step amounts to a replacement of the conventional on-shell
mass renormalization equation with the associated expression for the complex poles,
  \bq \label{eq:specialnote}
    m_i^2= 
    M_i^2 \left[ 
          1 + \frac{\gf \mws}{2\srt\, \pi^2} \text{Re} \Sigma_{i}^{(1)}(M_i^2)
          \right]
    \quad \Rightarrow \quad 
    m_i^2= 
    s_i \left[ 1+ \frac{\gf s_\ssW}{2\srt\, \pi^2} \Sigma_{i}^{(1)}(s_i)\right],
  \eq
where $\Sigma_{i}^{(1)}(M_i^2)$, with $i=W,Z$, denotes the $W$ ($Z$) one-loop 
self-energy, and complex poles are defined as
  \bq\label{eq:Cpoles}
    s_i= \mu_i \left(\mu_i - i \gamma_i \right),\qquad 
    \mu_i^2= M_i^2-\Gamma_i^2, \qquad
    \gamma_i= \Gamma_i \left(1 - \frac{\Gamma_i^2}{2 M_i^2 } \right).
  \eq
Here $M_i$ and $\Gamma_i$ are the canonical on-shell values for the mass
and the width of unstable gauge bosons. 

Note that, concerning the $W$ boson, the replacement of the real
part of the self-energy with the full expression has to be performed
also at the level of the Fermi-coupling renormalization equation, which becomes
  \bq\label{eq:fermi}
    g = 
    2 \left( \sqrt{2} \gf s_\ssW \right)^{1\slash 2} 
    \left[ 1- \frac{\gf s_\ssW}{4 \srt\, \pi^2} \Delta \right], \qquad 
    \Delta = 
    \Sigma_{\ssW}^{(1)}(0) - \Sigma_{\ssW}^{(1)}(s_\ssW) +
    6 + \frac{7 - 4 s_\theta^2}{2 s_\theta^2} \ln c_\theta^2.
  \eq
Here $g$ is the bare (or $\overline{MS}$-renormalized) weak-coupling constant, and 
the squared cosine of the weak-mixing angle $c^2_\theta= \cos^2 \theta$ 
($s^2_\theta= \sin^2 \theta= 1 - c^2_\theta$) is fixed by $c_\theta^2= \mu_\ssW^2 
\slash \mu_\ssZ^2$.

Here it is important to note that the cancellation mechanism between
two-loop diagrams and one-loop mass-renormalization terms generated by
the decomposition of \fig{4diags}a,b, mentioned commenting \eqn{eq:MCMsplit}, 
is a consequence of the introduction of \eqn{eq:specialnote} and does not take place in the
conventional on-shell renormalization framework for real masses.
This cancellation has a striking consequence: one-loop mass
renormalization contributes only to the remainder $A_{{\rm REM}}$, and
the steps summarized in \eqn{eq:specialnote} and \eqn{eq:fermi} are not required anymore.

\begin{figure}[ht]
\vspace{-0.4cm}
    \begin{center}
      \epsfig{file=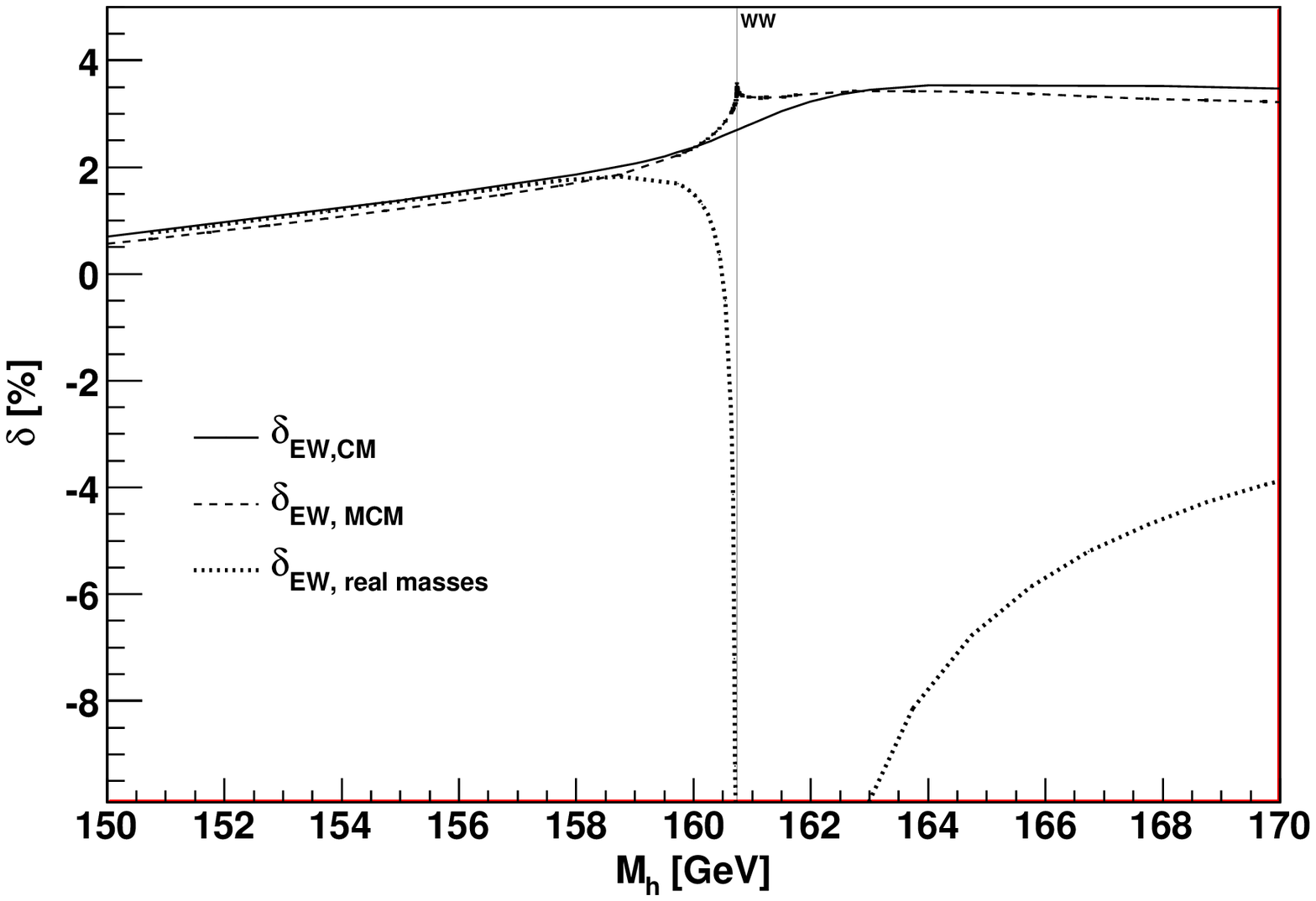, scale=0.60}
    \end{center}
    \vspace{-0.3cm}
    \caption[]{Percentage NLO electroweak corrections to the partial width for 
      $H\to \gamma \gamma$; here $\Gamma^{\rm NLO}= \Gamma^{\rm LO}(1+\delta_{\rm EW})$.
       MCM (CM) scheme is described in \sect{sec:MCMscheme} ($4.2$). 
       Setup is described in \sect{numa}.}
    \label{Delta}
\end{figure}

The MCM scheme allows for a straightforward removal of unphysical infinities:
real masses of unstable gauge bosons are traded for complex poles in
divergent terms, gauge-parameter invariance and WST identities are preserved
and the amplitude has a decent threshold behavior, as shown in \fig{Delta}
for the NLO electroweak corrections to the $H\to \gamma \gamma$  decay width.
Here the dotted curve represents the result obtained using conventional
on-shell masses for unstable gauge bosons as input data; the two-loop amplitude is 
artificially infinite at threshold and badly fails to approximate the correct
result above threshold, as a consequence of the severe $\beta_\ssW^{-1}$
behavior of \eqn{eq:MCMsplit} (enhanced above threshold by the fact that
the compensation illustrated in \fig{4diags}a,b does not occur in the pure
real-mass setup).

A nice feature of the MCM scheme (dashed curve) is its simplicity: as a consequence of the 
cancellation mechanism taking place among divergent contributions, this scheme 
does not require any analytic continuation of two-point functions connected to 
mass renormalization to the unphysical Riemann sheet, because the replacement 
indicated in \eqn{eq:specialnote} is not needed.
The MCM scheme, however, does not deal with cusps associated with the crossing of normal 
thresholds, as shown in \fig{Delta} for the $WW$ threshold.
\subsection{Complex-mass scheme
\label{sec:CMscheme}}
The large and artificial effects arising around normal thresholds in the MCM
scheme or in a scheme where the masses of unstable particles are kept real
(in this case the amplitude exactly diverges at threshold) are aesthetically 
unattractive. 
In addition, they represent a concrete problem in assessing the impact of two-loop
electroweak corrections on processes relevant for the LHC. An important example
is represented by Higgs production via gluon fusion: here, the large effect of 
NNLO QCD corrections naturally suggests to investigate possible sources of 
additional sizeable corrections. 

We have therefore undertaken the task of introducing the complete complex-mass 
scheme of Ref.~\cite{Denner:2005fg}, as explained for a two-loop calculation 
in Ref.~\cite{Actis:2006rc} (see also Ref.~\cite{Actis:2006ra}), for evaluating 
two-loop electroweak corrections to the Standard Model processes 
$H\to \gamma \gamma$ and
$g g \to H$. The procedure described in Section~\ref{sec:MCMscheme} for the 
divergent terms of \eqn{eq:MCMsplit} has been extended to the remainder 
$A_{{\rm REM}}$.
In particular, all two-loop diagrams have been computed with complex masses for 
the internal vector bosons relying on the techniques developed in 
Ref.~\cite{Passarino:2001wv,newLong}.

In the full CM setup, the real parts of the $W$ and $Z$ self-energies induced by 
one-loop renormalization of the masses and the couplings have to be traded for the 
associated complex expressions by means of \eqn{eq:specialnote} and \eqn{eq:fermi}.
However, we notice that:

i) for $H\to \gamma \gamma$, the $Z$ boson self-energy connected
to mass renormalization would arise only from the tree-level couplings of the 
photons, entailing an overall factor $g^2 s_\theta^2$ for the LO amplitude; in the 
CM scheme, in fact, $s_\theta^2$ is expressed through the ratio of the 
vector-boson masses.
However, because of our choice for the input-parameter set~\cite{Passarino:2007fp},
the factor $g^2 s_\theta^2$ is re-absorbed by introducing the fine-structure 
constant $\alpha$. In addition, as we said earlier, the electromagnetic coupling 
goes unrenormalized, once WFR factors for on-shell photons are included.

ii) for $g g\to H$, $Z$ mass renormalization clearly does not play any role. 
In addition, the tree-level coupling of the Higgs field to the top quark contains
a factor $g\slash \mw$; after combining the renormalization of the
weak-coupling constant $g$, related to the Fermi-coupling constant through 
\eqn{eq:fermi}, with mass renormalization for the $W$ boson, encoded in 
\eqn{eq:specialnote}, the $W$ self-energy evaluated at the complex pole drops out 
(see also Ref.~\cite{Degrassi:2004mx}).

As a result, for $gg\to H$ it is enough to replace the real masses for the $W$ and 
$Z$ bosons with their complex poles, as usual also in the couplings; for 
$H\to \gamma \gamma$, one has also to trade the real part of the $W$ self-energy 
for its full complex expression at the level of mass and coupling renormalization, 
via \eqn{eq:specialnote} and \eqn{eq:fermi}.
The effect for the $H\to \gamma \gamma$ decay mode, shown in \fig{Delta}
(solid curve), is a full smoothing of the unphysical cusp associated
with the $WW$ threshold; although numerical negligible, it provides
a benchmark for the $gg \to H$ study we will perform in Section~\ref{numa}.

The scheme can be easily extended to the fermionic sector replacing
also the top-quark real mass by its complex pole. Since the 
behavior associated with the $t\overline{t}$ threshold is $\beta_t\,$-protected, 
we do not purse this issue here.

The introduction of gauge-invariant complex poles for gauge bosons
leads to technical complications, due to the fact that one- and two-loop
integrals have to be computed with complex arguments. 
If only internal masses are complexified, the analytical continuation of
loop integrals does not pose any additional problem; after writing
the parametric representation of one- and two-loop diagrams, it is
easy to control that squared masses have semi-positive definite coefficients;
therefore the replacement $M^2 - i\,0 \to s_p = \mu^2 - i\,\mu\,\gamma$
is straightforward.

One-loop two-point functions arising in the reduction of the amplitude, 
instead, have to be carefully treated; here the external squared momentum 
can be complex and logarithms have to be extended to the second Riemann sheet. 
In general, the presence of complex momenta in two-loop diagrams demands an 
analytical continuation also for polylogarithms. 
In all cases, the correct analytical continuation is determined by the request 
that the value for a stable gauge boson should be smoothly approached when the 
coupling tends to zero.
This is achieved starting from a complex argument, $z= z_{\ssR} + i\,z_{\ssI}$,
defining ${\tilde z} = z_{\ssR} - i\,0$, and replacing ordinary logarithms
and polylogarithms with
  \bq
  {\overline\ln} \lpar z\,;\,{\tilde z} \rpar = 
  \ln z - 2\,i\,\pi\,\theta \lpar - z_{\ssR} \rpar,
  \qquad
  {\overline{\text{Li}}}_2 \lpar z\,;\,{\tilde z} \rpar = 
  \li{2}{z} - 2\,\,i\,\pi\,\ln z_{\ssR}\,\theta\lpar z_{\ssR} - 1 \rpar .
  \eq
\subsection{External unstable particles}
From a formal perspective, external unstable particles should not appear
in any computation, since they cannot be included in the asymptotic
states forming the bases of the Hilbert space. Concerning Higgs physics, 
however, available calculations deal with an external on-shell Higgs boson 
and do not perform the ultimate step of introducing a complex pole and the 
associated residue for the decaying Higgs~\cite{Kniehl:2001ch}.

For the production process $g g \to H$, we have verified that there are no 
practical problems associated with gauge-parameter invariance and WST identities
once we deal with an on-shell Higgs.
For the $H \to \gamma \gamma$ decay, there is a LO contribution containing the 
bare Higgs mass, represented by a charged Higgs-Kibble one-loop triangle diagram. 
Standard mass renormalization introduces the on-shell Higgs mass, through the real 
part of the one-loop two-point self-energy, and leads to a violation of the WST 
identities above the $WW\,$ threshold. 
In both aforementioned MCM and CM schemes the real part of the Higgs self-energy 
stemming from mass renormalization is traded for the complex expression, even if 
the external Higgs boson is assumed to be an on-shell particle.
\section{Numerical effects for Higgs production}
\label{numa}
The dominant production mechanism of the Standard Model Higgs boson at the 
LHC is the gluon fusion process, $gg \to H$: in this section we discuss
the numerical impact at the two-particle vector-boson thresholds 
of the two-loop electroweak corrections in the framework
of the MCM and CM schemes.

In the computation we have set light-fermion masses to zero and 
introduced the $W$ and $Z$ boson complex poles by means of \eqn{eq:Cpoles}.
As input parameters we have used the following values taken from Ref.~\cite{PDG}:
\bqa
\begin{array}{ll}
\mw = 80.398\,\GeV,  \;\; & \;\; 
\mz = 91.1876\,\GeV, \;\; \\
\Gamma_{\ssZ} = 2.4952\,\GeV, \;\; & \; \;
\gf = 1.16637\,\times\,10^{-5}\,\GeV^{-2}.
\end{array}
\eqa
For the mass of the top quark, we have used $M_t=170.9\,\GeV$~\cite{Top};
for the width of the $W$ boson, we have chosen the value
$ \Gamma_{\ssW} = 2.093\,\GeV$, predicted by the Standard Model
with electroweak and QCD corrections at one loop.

\begin{figure}[ht]
\begin{center}
\includegraphics[scale=0.55]{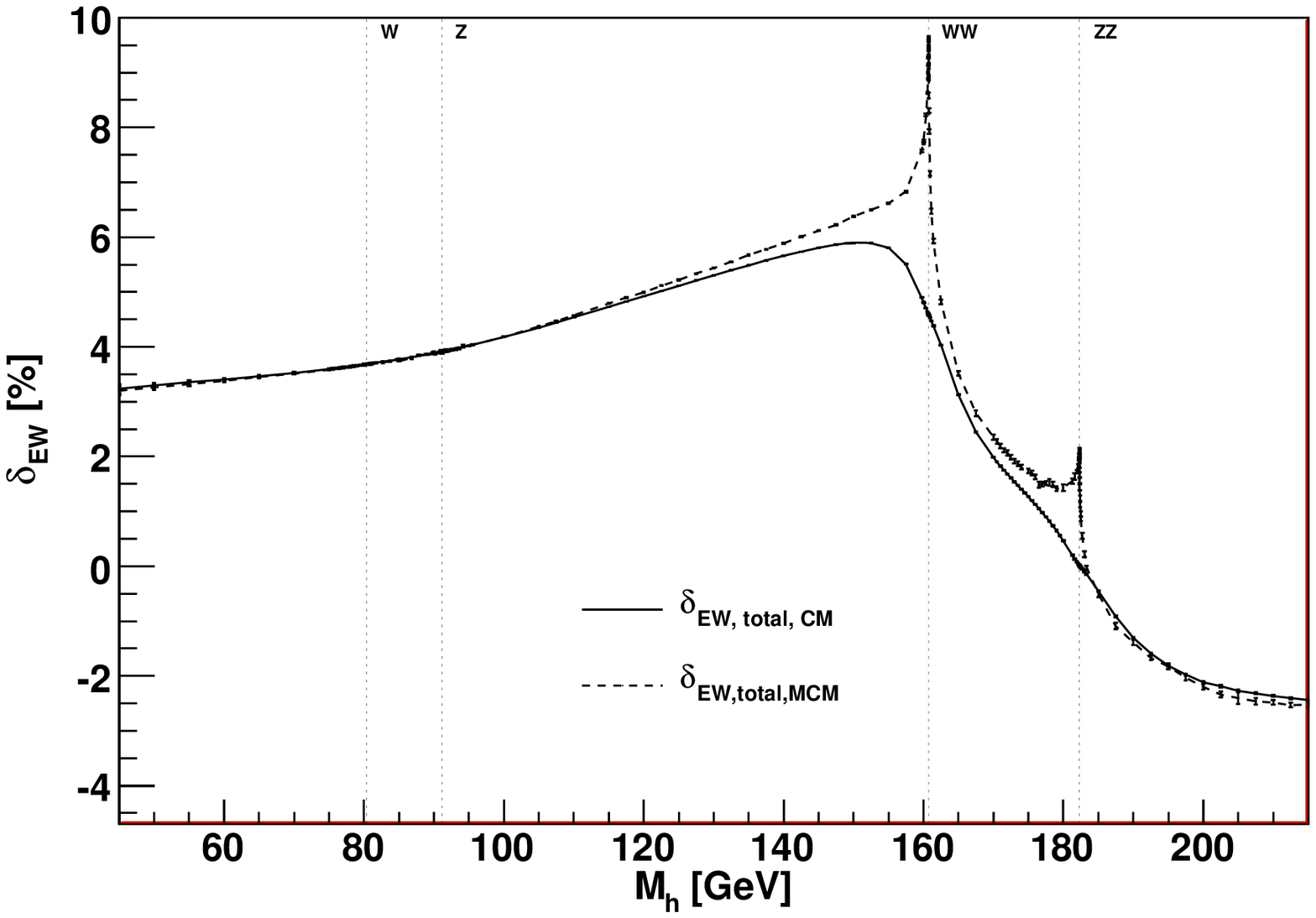}
\vspace{-0.3cm}
\caption[]{ The two-loop electroweak percentage
  corrections for the total partonic cross section
  $\sigma( g \, g \to H)$.
  MCM (CM) scheme is described in \sect{sec:MCMscheme} ($4.2$).
  Setup of \sect{numa}.}
\label{fig:dEWHgluglu}
\end{center}
\end{figure}

Our results for $\delta_{\EW}$, defined through $
\sigma^{\rm EW}= \sigma^{\rm LO}(1+\delta_{\EW})$, are shown 
in \fig{fig:dEWHgluglu}, where we include the complete corrections, 
comprehensive of light- and top-quark contributions, comparing
MCM and CM scheme. The corresponding numerical results are given in
\tabn{tab:anco}. 
\renewcommand{\arraystretch}{1.1}
\begin{table}[ht]
\vspace{0.3cm}
  \begin{center}
    \begin{tabular}{|r|r|r||r|r|r||r|r|r|}
      \hline
      $M_H$ & $\delta_{\EW}^{ \rm MCM}$  & $\delta_{\EW}^{\rm CM}$ & 
      $M_H$ & $\delta_{\EW}^{ \rm MCM}$  & $\delta_{\EW}^{\rm CM}$ & 
      $M_H$ & $\delta_{\EW}^{ \rm MCM}$  & $\delta_{\EW}^{\rm CM}$ \\
      \hline
      \hline
      140.0 &5.88 & 5.66 &  162.5 & 4.82 & 4.03 & 180.0 & 1.43    & 0.47 \\
      145.0 &6.12 & 5.80 &  165.0 & 3.52 & 3.13 & 182.5 & 0.96    & $-$ 0.02 \\
      150.0 &6.38 & 5.90 &  167.5 & 2.79 & 2.45 & 185.0 & $-$ 0.50& $-$ 0.46 \\
      152.5 &6.50 & 5.89 &  170.0 & 2.35 & 1.99 & 187.5 & $-$ 1.09& $-$ 0.91 \\
      155.0 &6.62 & 5.81 &  172.5 & 1.99 & 1.61 & 190.0 & $-$ 1.39& $-$ 1.31 \\
      157.5 &6.83 & 5.51 &  175.0 & 1.74 & 1.27 & 195.0 & $-$ 1.82& $-$ 1.82 \\
      160.0 &7.72 & 4.82 &  177.5 & 1.51 & 0.90 & 200.0 & $-$ 2.20& $-$ 2.11 \\
      \hline 
    \end{tabular}
  \end{center}
  \vspace{-0.2cm}
  \caption[]{Percentage electroweak corrections
    in MCM (\sect{sec:MCMscheme}) and CM (\sect{sec:CMscheme}) schemes as a 
    function of the Higgs mass in GeV ($M_t=170.9\,\GeV$).}
  \label{tab:anco}
\end{table}

\begin{figure}[ht]
\vspace{-0.2cm}
\begin{center}
\includegraphics[scale=0.55]{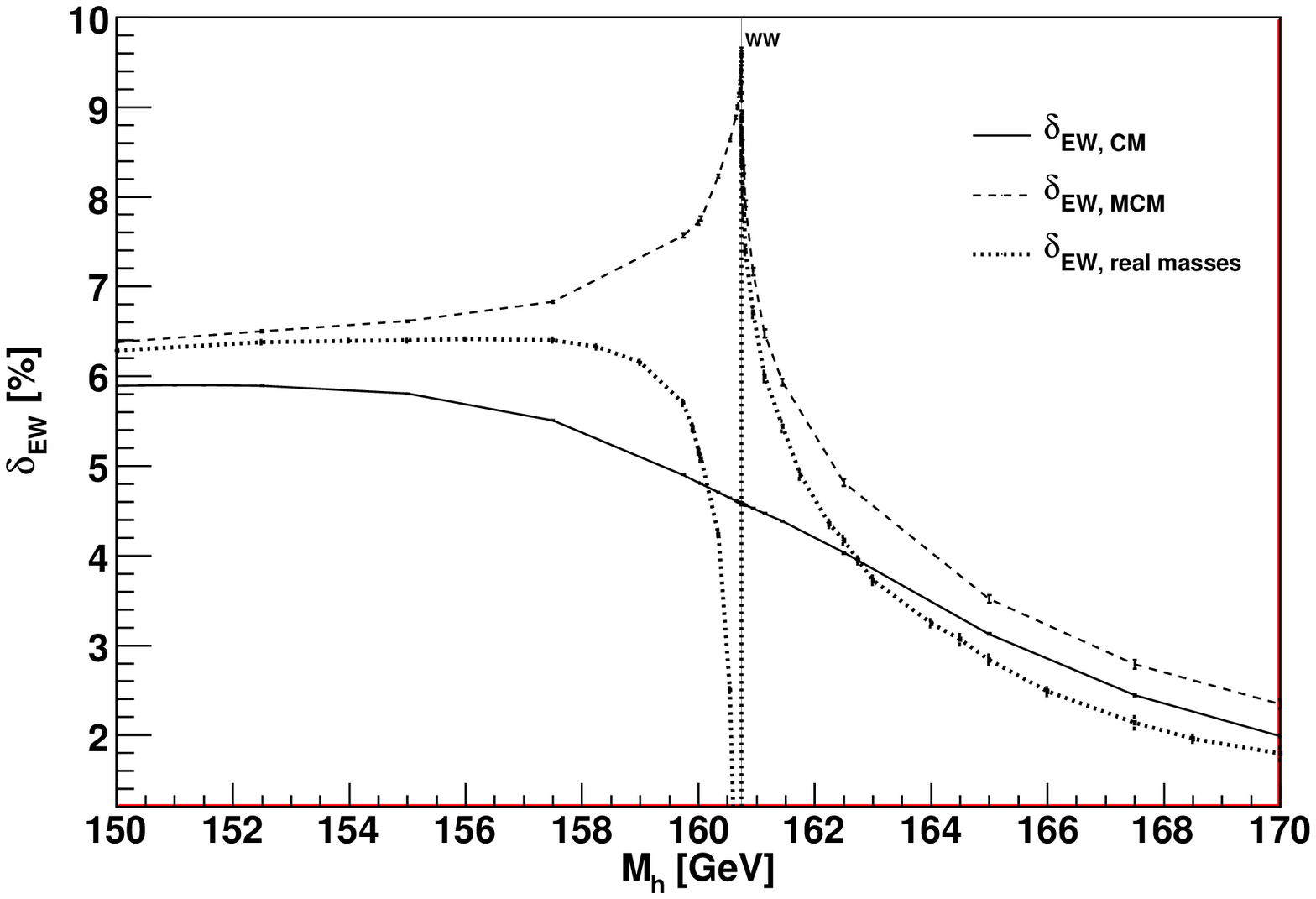}
\vspace{-0.3cm}
\caption[]{ Same as \fig{fig:dEWHgluglu}, around the $WW$ threshold.
MCM (CM) scheme is described in \sect{sec:MCMscheme} ($4.2$).
Setup of \sect{numa}.}
\label{fig:glus}
\end{center}
\vspace{-0.4cm}
\end{figure}

In \fig{fig:glus} we show the details of the region
around the $WW$ threshold, including the result obtained using
purely real masses.

The numerical relevance of NLO electroweak corrections around the vector-boson
thresholds depends crucially on the implementation of the renormalization
scheme: the relative corrections in the MCM scheme reach about $10 \%$ ($2 \%$)
at the $WW$ ($ZZ$) threshold; in the CM scheme, they amount to $5 \%$ at the
$WW$ threshold and vanish at the $ZZ$ one.
\section{Conclusions}
In this paper we have considered the extension of the complex-mass scheme to
two-loop multi-scale calculations. After discussing the general setup we have given
numerical results for the two-loop percentage corrections to the
$H \to \gamma \gamma$ and $g g \to H$ processes in the Standard Model around
the vector-boson thresholds. 

We have compared a minimal implementation of the complex-mass scheme
and the complete one. For Higgs masses close to the $WW$ and $ZZ$ thresholds,
NLO electroweak corrections can be considered under control only
after a full implementation of the complex-mass scheme.

The electroweak scaling factor for the cross section does not exceed the 
$+6\, \%$ level in the range $100\,\GeV < \mh < 200\,\GeV$; incongruent 
$+10\, \%$ effects around thresholds are avoided, as a
consequence of the complex-mass scheme employed.
\Acknowledgments
Feynman diagrams have been drawn with the packages 
{\sc Axodraw}~\cite{Vermaseren:1994je} and 
{\sc Jaxo\-draw}~\cite{Binosi:2003yf}.
The calculations performed in this paper have been performed
with {\sc FORM}~\cite{Vermaseren:2000nd}.

\end{document}